\documentclass[12pt,a4paper]{article}
\usepackage[latin1]{inputenc}
\usepackage[T1]{fontenc}
\usepackage{amsmath}
\usepackage{amsfonts}
\usepackage{amssymb}
\usepackage[english]{babel}
\title{Quantum effects in $QED_{3+1}$ with an extra compactified dimension}
\author{C.~Ccapa Ttira$^{a}$, C.~D.~Fosco$^{a}$,\\
A.~P.~C.~Malbouisson$^{b}$\\ and I.~Roditi$^{b}$\\ {\normalsize\it
$^a$Centro At\'omico Bariloche and Instituto Balseiro}\\
{\normalsize\it Comisi\'on Nacional de Energ\'\i a At\'omica}\\
{\normalsize\it 8400 Bariloche, Argentina.}\\ {\normalsize\it
$^b$Centro Brasileiro de Pesquisas F\'isicas - CBPF/MCT}\\
{\normalsize\it Rua Dr. Xavier Sigaud, 150, 22290-180 Rio de Janeiro,
RJ, Brazil}}
\begin{document}
\maketitle
\begin{abstract}
\noindent We evaluate one-loop effects for $QED_{3+1}$ in the presence of
an extra compactified dimension, in a non-trivial vacuum for the gauge
field, such that a non-vanishing magnetic flux is encircled along the extra
dimension. We obtain the vacuum polarization tensor and evaluate the exact
parity breaking term.
\end{abstract}



Quantum field theory models with compactified dimensions have been proposed
to investigate many different physical phenomena. These range from finite
size effects in critical phenomena~\cite{Brezin:1985xx}, to particle
physics models where, since the pioneering works of Kaluza and
Klein~\cite{KK}, extra dimensions have played an important role to search
for methods to unify the fundamental interactions and to 
describe a wide range of physical situations; to name a few:
alternative solutions to the hierarchy problem~\cite{dvali,sundrum},
physics at the TeV scale~\cite{antoniadis1}, and particle physics models
where the extra dimensions play an important role to implement the
unification of fundamental interactions (see, for example,~\cite{dienes,Lawrence,cvetic}
and references therein) . 

An interesting yet simple example of this, corresponds to the
compactification of one dimension in an ${\mathbb R}^D$ Euclidean spacetime, such
that the topology of the resulting manifold ${\mathcal M}$ is that of
\mbox{${\mathbb R}^{D-1} \times {\mathcal S}_ 1$}, i.e., `circular
compactification'.  Although the interesting features that emerge in this
situation have been studied using several different techniques in the
literature, one can take advantage of a (formal) common feature; indeed,
they share many properties with the imaginary-time formulation of quantum
field theory at finite temperature~\cite{3mats1,3ume4}.  This allows one,
for example, to take advantage of the many well-known methods and results
developed in this context, such as Feynman diagrams and renormalization
techniques, to import them to the case under consideration. 

From a topological point of view, the imaginary-time formalism in a
$D$-dimensional space-time at finite temperature is equivalent to a
path-integral evaluated on the manifold \mbox{${\mathbb R}^{D-1} \times
{\mathcal S}_ 1$}, where $S_{1}$ is a circle of circumference $\beta =1/T.$
Since in Euclidean path integral (imaginary) time and spatial coordinates
are on the same footing, the Matsubara prescription can be thought of as a
mechanism to deal with spatial compactifications where the compactification
length $L$ plays the role of $\beta$~\cite{ford1}.  This has been developed
by considering both simply or non simply-connected $D$-dimensional
manifolds with a topology ${\mathbb R}^{D}_{d}={\mathbb R}^{D-d}\times
S_{l_{1}}\times {\mathcal S}_{l_{2}}\cdots \times {\mathcal S}_{l_{d}}$,
with $l_{1}$ corresponding to the inverse temperature and $l_{2}\,,\cdots
l_{d}$  to the compactification of $d-1$ spatial
dimensions~\footnote{This case has been considered, within the context of spontaneous 
symmetry breaking, in~\cite{Ademir}.}. 

For just one compactified dimension (imaginary time or a spatial
dimension) the Feynman rules are modified, the most characteristic new
feature is the Matsubara prescription for momentum integrals,  
\begin{equation}
\int \frac{dk_s}{2\pi }\rightarrow \frac 1{\xi}\sum_{n=-\infty }^{+\infty
}\;;\;\;\;k_i\rightarrow \frac{2n\pi }{\xi}\;,
\label{Matsubara1}
\end{equation}
where $k_s$ amounts to the momentum component corresponding to the compactified
dimension, while $\xi$ equals $\beta$ or $L$, for the finite temperature and
compactified spatial dimension cases, respectively.

Some of these ideas have recently attracted renewed interest, as a way to
investigate the electroweak transition and baryogenesis. For instance, a
recent  investigation of the electroweak phase transition has been
presented in~\cite{Panico}, in the context of a $4+1$-dimensional theory
with a compactified dimension. This model involves charged scalar bosons
coupled to a gauge field, and makes use of the Hosotani
mechanism~\cite{Hosotani}.  Taking the order parameter to be proportional
to the non-vanishing component of the gauge field along the compactified
coordinate, the authors find a first-order transition with a strength
inversely proportional to the Higgs mass. 

Within the previous general framework, we here take a slightly different route, 
to investigate one-loop effects for $QED_{3+1}$ with an extra compactified dimension, 
in a non-trivial vacuum for the gauge field, with the non-vanishing component of the gauge field
lying also along the extra dimension.  

The system we shall deal with may be conveniently defined in terms of an 
Euclidean action, $S$, which has the structure:
\begin{equation}\label{eq:defs0}
S({\mathcal A};{\bar\Psi},\Psi) \;=\; S_g({\mathcal A}) 
\,+\, S_f({\mathcal A};{\bar\Psi},\Psi)\;, 
\end{equation}
where $S_g$ and $S_f$ denote the $U(1)$ gauge field and fermionic actions,
respectively. The former is assumed to have a standard Maxwell form, namely:
\begin{equation}\label{eq:defsg}
S_g({\mathcal A}) \;=\; \frac{1}{4} \, \int d^5x \, {\mathcal
F}_{\alpha\beta}  {\mathcal F}_{\alpha\beta} \;,
\end{equation}
with ${\mathcal F}_{\alpha\beta}\equiv\partial_\alpha {\mathcal A}_\beta -
\partial_\beta {\mathcal A}_\alpha$, where we adopted the convention that
indices from the beginning of the Greek alphabet ($\alpha$, $\beta$,
\ldots) label all the coordinates of the spacetime manifold, and therefore
run from $0$ to $4$.  Since we will be specially interested in the model as
it is seen from a $3+1$ dimensional point of view, we shall also use
another convention: indices from the middle of the Greek alphabet ($\mu$,
$\nu$, \ldots) are reserved for the $3+1$-dimensional spacetime
coordinates while, when this notation is used, the extra dimension
coordinate shall be denoted by $s$. Then:
\begin{equation}
\alpha \,=\, 0,\,1,\,2,\,3,\,4 \;,\;\;\;
\mu \,=\, 0,\,1,\,2,\,3\;,\;\;\;
d^5 x \;\equiv\; d^{3+1}x \,dx_4 \;=\; d^{3+1}x \, ds \;,
\end{equation}
and $x$ will be assumed to denote the $3+1$ coordinates $x_\mu$,
unless explicit indication on the contrary.  The extra dimension is assumed
to be compactified with a radius $R$, so that $s \sim s + L$, $L = 2 \pi
R$.

On the other hand, the Dirac action, $S_f$, is given by
\begin{equation}
 S_f({\bar\Psi},\Psi;{\mathcal A}) \;=\; \int d^{3+1}x \, ds \; {\bar\Psi}(x,s) 
\big( {\mathcal D} + m \big) \Psi(x,s)
\end{equation} 
where ${\mathcal D}$ is the $4+1$ dimensional Dirac operator, \mbox{${\mathcal
D}= \gamma_\alpha D_\alpha$}. The covariant derivative 
\mbox{$D_\alpha \equiv \partial_\alpha + i g {\mathcal A}_\alpha$} 
includes a coupling constant $g$ with the dimensions of $({\rm
mass})^{-\frac{1}{2}}$. For Dirac's $\gamma$-matrices, we assume that
$\gamma_s \equiv \gamma_5$, where the latter is the $\gamma_5$ matrix for
the $3+1$ world. 

To proceed, we discuss now the mode expansion and its relation to gauge
invariance. To that end, we follow~\cite{zinn}, where this issue is
discussed at length, albeit in the finite temperature theory context, in the
Matsubara formulation of thermal field theory. Due to the formal analogy  
with this situation, a quite straightforward procedure allows us to adapt 
the results derived there to our case. The necessary changes that 
follow from the fact that our compactified dimension is spatial rather 
than temporal are taken into account by using (\ref{Matsubara1}). 
In that analogy, the length $L$ plays the same role of 
the inverse temperature in \cite{zinn}: $L \sim \beta$, $\beta = T^{-1}$.

What follows is a brief review of some of those properties (the ones which
are relevant to our study), adapted to our case and conventions. To begin
with, the gauge field configuration ${\mathcal A}_\alpha(x,s)$ may be
decomposed into its zero ($A_\alpha$) and non-zero ($Q_\alpha$) mode
components:
\begin{equation}
{\mathcal A}_\alpha(x,s) \;=\; L^{-\frac{1}{2}} \, A_\alpha(x) \,+\,
Q_\alpha(x,s) \;,
\end{equation} 
where the two terms in this decomposition may be defined by:
\begin{equation}
A_\alpha(x) \;=\;  L^{-\frac{1}{2}} \, \int_0^L ds \, 
{\mathcal A}_\alpha (x,s) \;,
\end{equation}
and 
\begin{equation}
Q_\alpha(x,s) \;=\; {\mathcal A}_\alpha(x,s) \,-\, L^{-\frac{1}{2}} \,
A_\alpha(x) \;, 
\end{equation} 
so that $\int_0^L ds \, Q_\alpha(x,s)\,=\, 0$. An $L^{-\frac{1}{2}}$ factor
has been included  in the zero mode term in order to make this field have
the usual mass dimensions in $3+1$ spacetime dimensions; this property will
become useful after dimensional reduction. 

The decomposition above finds a natural interpretation when one considers the
Fourier expansion of the gauge field along the extra dimension:
\begin{equation}
{\mathcal A}_\alpha (x,s) \;=\; L^{-\frac{1}{2}} \, 
\sum_{n=-\infty}^{\infty}\, e^{ i \omega_n s} \,\widetilde{\mathcal A}_\alpha(x,n) \;,
\end{equation}
with $\omega_n \equiv \frac{2 \pi n}{L}$, where one identifies:
\begin{equation}
A_\alpha(x) \,=\, \widetilde{\mathcal A}_\alpha(x,0) \;\;,\;\;\;\; 
Q_\alpha(x,s) \,=\, L^{-\frac{1}{2}} \, 
\sum_{n \neq 0}\, e^{ i \omega_n s} \,\widetilde{\mathcal A}_\alpha(x,n) \;.
\end{equation}

Then we dimensionally reduce the theory, what, for the gauge field action,
amounts to keeping just the zero mode component of the gauge field. Thus:
\begin{equation}
S_g({\mathcal A}) \; \to \; S_g (A) \;=\; S_g(A_\mu, \,A_s) \;, 
\end{equation}
where:
\begin{equation}
S_g(A_\mu, A_s) \;=\; \int d^{3+1}x \, \Big[
\frac{1}{2} \partial_\mu A_s \partial_\mu A_s 
\,+\, \frac{1}{4} F_{\mu\nu}(A) F_{\mu\nu}(A) \Big] \;,
\end{equation}
with $F_{\mu\nu}(A) \equiv \partial_\nu A_\nu - \partial_\nu A_\mu$.

Regarding the fermionic action $S_f$, the reduction amounts to:
\begin{equation}
S_f({\mathcal A}; {\bar\Psi}, \Psi ) \; \to \; 
S_f(A_\mu,A_s;{\bar\Psi},\Psi) \;. 
\end{equation}
The fermionic field is not dimensionally reduced by the simple reason that,
in the calculation of the effective gauge field action, its only contribution
comes from the fermion loop. That loop may be represented as a series of
$3+1$ loops, each one with a different mass. Although the contributions of
heavier modes may be relatively suppressed, the very fact that there is an
infinite number of them forbids us to truncate that series (even if there
were a zero mode).

Thus, the following explicit expression for the fermionic action shall be
used after dimensional reduction:
\begin{equation}
S_f \;=\; \int d^{3+1} x \, \int_0^L ds \; {\bar\Psi}(x,s) 
\big( \not \!\! D + \gamma_s D_s  + m \big)  \Psi(x,s)
\end{equation} 
where
\begin{equation}
\not \!\! D \,=\, \gamma_\mu (\partial_\mu + i e A_\mu )
\;\; D_s \,=\, \partial_s + i e A_s  \;.
\end{equation}
We have introduced a new, dimensionless coupling constant $e \equiv g
L^{-\frac{1}{2}}$, which shall play the role of the electric charge in $3+1$
dimensions.

As explained in~\cite{zinn}, when considering the form of the gauge
transformations in terms of the decomposition into zero and non-zero modes,
one finds that it $A_\mu$ transforms as a standard gauge field (in $3+1$ dimensions):
\begin{equation}
\delta A_\mu (x) \,=\, \partial_\mu \alpha (x)
\end{equation}
while its extra dimensional component $A_s$, a scalar from the
$3+1$-dimensional point of view, is shifted by a constant:
\begin{equation}
\delta A_s (x) \,=\,\Omega \;.
\end{equation}
The constant $\Omega$ has to be of the form~\mbox{$\Omega = \frac{2 \pi
k}{L e}$}, where $k$ is an integer, since the gauge field is
coupled to a (charged) fermionic field, whose transformation law under
simultaneous action of the previous gauge transformations is:
\begin{eqnarray}
\Psi (x,s) &\to&  e^{-i e [\alpha(x) + \Omega s ]} \, \Psi (x,s) \nonumber\\
\bar{\Psi} (x,s) &\to&  e^{ i e [\alpha(x) + \Omega s ]} \,
\bar{\Psi} (x,s) \;.
\end{eqnarray}  

We now define the part of the effective action that only
depends on the (dimensionally reduced) gauge field, $\Gamma (A)$, 
\begin{equation}
\Gamma (A) \; \equiv \; \Gamma (A; {\bar\Psi}, \Psi) \Big|_{{\bar\Psi}=
\Psi = 0} \;,
\end{equation}
where $\Gamma (A; {\bar\Psi}, \Psi)$ is the full effective action. 
The functional $\Gamma(A)$ allows one to derive $1PI$ functions containing 
only $A_\mu$, $A_s$ external lines. The former have an immediate $3+1$
dimensional interpretation, while the latter shall be assumed to have
a constant (but otherwise arbitrary) value, which is determined by a
condition which is external to the model.

On the other hand, at the one-loop order, the only non-trivial term comes 
from the fermionic loop:
\begin{equation}
\Gamma(A) \;=\; \Gamma^{(0)}(A) \,+\,\Gamma^{(1)}(A) \;+\;\ldots  
\end{equation}
where $\Gamma^{(0)}(A) = S_g(A)$ and 
\begin{equation}
e^{-\Gamma^{(1)}(A)} \;=\; \int {\mathcal D}\Psi {\mathcal D}{\bar\Psi} e^{-
S_f(A; {\bar\Psi},\Psi) } \;. 
\end{equation}

We shall focus on the effective action for the gauge field components
$A_\mu$ that have a direct physical interpretation from a $3+1$-dimensional
perspective. Regarding the scalar component, $A_s$, as we have said above, it 
will be assumed to yield a non-vanishing flux:
\begin{equation}
e \int_0^L ds A_s \;=\; \phi 
\end{equation}
where $\phi$ is a constant.  This condition may be conveniently solved by
means of a constant $A_s$: 
\begin{equation}
A_s \;=\; \frac{\phi}{e L} \;,
\end{equation}
which is the gauge fixing that we shall assume. Note that, since the gauge
transformations shift $A_s$ by an integer multiple of $\frac{2\pi}{e L}$,
we may fix the value of $\phi$ to the fundamental region:
\begin{equation}
0 \leq \phi < 2 \pi  \;,
\end{equation}
which we shall assume in what follows.

We then proceed to Fourier expand the fermionic fields along the $s$
coordinate:
\begin{eqnarray}
\Psi(x,s) &=& L^{-\frac{1}{2}} \,\sum_{n=-\infty}^\infty e^{ i \omega_n s} \psi_n(x)
\nonumber\\
{\bar\Psi}(x,s) &=& L^{-\frac{1}{2}} \,\sum_{n=-\infty}^\infty e^{ - i \omega_n s}
{\bar\psi}_n(x) \;,
\end{eqnarray}
and insert this into the functional expression for $\Gamma^{(1)}(A)$, to
obtain:
\begin{equation}\label{eq:sfexpand}
S_f \;=\; \sum_{n=-\infty}^{n=+\infty} 
\int d^{3+1}x\,\bar{\psi}_n(x) \big(\not \!\! D + i \gamma_s (\omega_n +
\frac{\phi}{L} ) + m \big) \psi_n(x) \;.
\end{equation}
Under the same expansion, the fermionic measure factorizes:
\begin{equation}
\mathcal{D}\Psi \mathcal{D}\bar{\Psi}\;=\;
\prod_{n=-\infty}^{n=+\infty} \mathcal{D}\psi_n(x) \mathcal{D}\bar{\psi}_n(x),
\end{equation}
and, finally, the Euclidean action corresponding to each mode $n$ may be
equivalently written as follows
$$
\int d^{3+1}x\,\bar{\psi}_n(x) \big(\not \!\! D + i \gamma_s (\omega_n +
\frac{\phi}{L} ) + m \big) \psi_n(x) 
$$
\begin{equation}
=\; \int d^{3+1}x \,\,\bar{\psi}_n(x)(\not \!\! D + M_n \,
e^{-i \varphi_n \gamma_5})\psi_n(x)
\end{equation}
with
\begin{equation}
 M_n \equiv \sqrt{m^2+(\omega_n+\phi/L)^2}\;,\;\;\; 
\varphi_n = {\rm arctan}(\frac{\omega_n+\phi/L}{m}) \;.
\end{equation}
The existence of a $\gamma_5$ term means that parity symmetry will generally be
broken; to study that phenomenon more clearly, we perform  a 
change in the fermionic variables that gets rid of the dependence in $\gamma_5$,
\begin{equation}
\psi_n(x)\to e^{-i\gamma_5 \phi_n/2}\psi_n(x)\,,\,\,\,\,\bar{\psi}_n(x) 
\to \bar{\psi}_n(x)e^{-i\gamma_5 \phi_n/2} \;,
\end{equation}
after which the mode labelled by $n$ has the action:
\begin{equation}
\int d^{3+1}x \,\,\bar{\psi}_n(x)(\not \!\! D + M_n \,
e^{-i \varphi_n \gamma_5})\psi_n(x) \;.
\end{equation}
This chiral rotation in the $3+1$ Euclidean fermionic variables induces,
however an anomalous Jacobian ${\mathcal J}_n$ for each mode. 
Then, $\Gamma^{(1)}$ may be written as follows:
\begin{equation}
e^{-\Gamma^{(1)}(A)} \;=\; \prod_{n=-\infty}^{+\infty}\Big[  
{\mathcal J}_n \; e^{-\Gamma^{(1)}_{3+1}(A,M_n)} \Big]\;,
\end{equation}
where 
\begin{equation}\label{eq:defj}
{\mathcal J}_n \;=\; \exp \big(\frac{i e^2}{16 \pi^2} \, 
\phi_n \int d^{3+1}x {\tilde F}_{\mu\nu} F_{\mu\nu} \big)  \;,
\end{equation}
with ${\tilde F}_{\mu\nu}=\frac{1}{2} \epsilon_{\mu\nu\rho\lambda}
F_{\rho\lambda}$,
and $\Gamma^{(1)}_{3+1}(A,M_n)$ is the one-loop fermionic contribution to the
effective action, for a fermion whose mass is $M_n$, in $3+1$ dimensions. 
Of course, it may be expressed as a fermionic determinant:
\begin{equation}
e^{-\Gamma^{(1)}_{3+1}(A,M_n)} \;=\; \det(\not \!\! D + M_n) \;.
\end{equation}
Then, we arrive to a general expression for the one loop effective action,
\begin{equation}
\Gamma^{(1)}(A) \;=\;\Gamma^{(1)}_e (A) \,+\, \Gamma^{(1)}_o (A) 
\end{equation}
where the $e$ and $o$ subscripts stand for the even an odd components
(regarding parity transformations) and are given by
\begin{equation}
\Gamma^{(1)}_e(A) \;=\; \sum_{n=-\infty}^{\infty} \Gamma^{(1)}_{3+1}(A,M_n) 
\end{equation}
and 
\begin{equation}\label{eq:sumj}
\Gamma^{(1)}_o(A) \;=\; - \sum_{n=-\infty}^{\infty} \ln {\mathcal J}_n \;,
\end{equation}
respectively.

The parity conserving part of the effective action may be obtained by
performing the sum of the required $QED_{3+1}$ object, with an $n$-dependent
mass, $M_n$.  We shall focus on that part of $\Gamma^{(1)}_e$ that
contributes to the vacuum polarization tensor for the $A_\mu$ gauge field components.
Since we are not interested in response functions which involve the $s$
component of the currents, it is useful to define:
\begin{equation}
\Gamma^{(1)}_e (A_\mu) \;\equiv\;\Gamma^{(1)}_e (A_\mu,A_s) \,-\,
\Gamma^{(1)}_e (0,A_s) \;.   
\end{equation}
Note that $\Gamma^{(1)}_e (0,A_s)\equiv \Gamma_s(A_s)$ does not contribute
to response functions involving $A_\mu$, although it can be used to study
the fermion-loop corrections to an $A_s$ effective potential. The explicit
form of this function is~\cite{zinn}:
\begin{equation}
 \Gamma_s(A_s) \,=\,- 2 L \int d^{3+1}x \, \int \frac{d^4k}{(2\pi)^4} 
\ln \big[ \cosh(L k) \,+\, \cos\phi \big] \;. 
\end{equation}

The vacuum polarization tensor $\Pi_{\mu\nu}$  is obtained from the quadratic 
term in a functional expansion in the gauge field:
\begin{equation}
\Gamma^{(1)}_e(A_\mu) \;=\; \frac{1}{2} \int d^{3+1}x \int d^{3+1}y 
A_\mu(x) \Pi_{\mu \nu}(x,y) A_\nu(y) \;+\; \ldots 
\end{equation}

It is then sufficient to resort to the analogous expansion for the $3+1$ 
dimensional effective action,
\begin{equation}
\Gamma_{3+1}^{(1)}(A,M_n) \,=\; \frac{1}{2} \int
d^{3+1}x \int d^{3+1}y A_\mu(x) \Pi_{\mu \nu}^{(n)}(x,y) A_\nu(y)
\;+\; \ldots 
\end{equation}
(which is even) so that the vacuum polarization receives contributions from
all the modes:
\begin{equation}\label{eq:pisum}
\Pi^e_{\mu \nu} \;=\;\sum_n \Pi_{\mu \nu}^{(n)}\,,
\end{equation}
where $\Pi_{\mu \nu}^{(n)}=\Pi^{(n)}(k^2) \; \delta^T_{\mu \nu}(k)$, with:
\begin{equation}
\Pi^{(n)}(k^2) \;=\;
\frac{2\,e^2}{\pi} \int_0^1 d\beta\,\,\beta(1-\beta)\,\,\ln 
\left[1+\beta(1-\beta)\frac{k^2}{M_n^2}\right]\;,
\end{equation}
the renormalized scalar part of the vacuum polarization tensor,
and the transverse projector is defined by 
\mbox{$\delta^T_{\mu \nu}(k) \equiv  \delta_{\mu\nu} - k_\mu k_\nu/k^2$}.

The sum in (\ref{eq:pisum}) may be evaluated using standard finite
temperature techniques; this yields the result $\Pi^e_{\mu\nu} = \Pi_e
\, \delta^\perp_{\mu\nu}$, with: 
$$
\Pi_e(k^2) \;=\; \frac{2\,e^2}{\pi} \int_0^1 d\beta\; \beta(1-\beta)
$$
\begin{equation}
\times \ln \Big\{ \cos(\phi)- \cosh \big[ m L 
\sqrt{1+\beta(1-\beta)\frac{k^2}{m^2}} \big] \Big\} \;.
\end{equation}

We see that, as a consequence of the sum over modes, the expression above
does not satisfy the renormalization condition that follows by imposing the
validity of Coulomb's law at long distances. 
However, it is quite straightforward to
impose it now, since we only need to perform a finite renormalization, to
obtain a properly renormalized function $\Pi_R$, whose explicit form is:
\begin{eqnarray}\label{eq:piren1}
\Pi^e_R (k^2) &=& \Pi_e (k^2) \,-\, \Pi_e(0) \nonumber\\
&=& \frac{2\,e^2}{\pi} \int_0^1 d\beta\; \beta(1-\beta)
\; \ln [ 1 + F(k^2) ]\;,
\end{eqnarray}
with
\begin{equation}\label{eq:piren2}
F(k^2) \;=\; \frac{\cosh \big[m L
\sqrt{1+\beta(1-\beta)\frac{k^2}{m^2}}\big] -
\cosh(m  L)}{\cosh (m L) - \cos(\phi)}\;.
\end{equation}
It is interesting to note that, even though the theory is $5$ dimensional,
the vacuum polarization tensor requires, to be renormalized, just one
renormalization condition, as in $4$ dimensions. Indeed, the superficial
degree of divergence, $\delta (\gamma)$, for an $1PI$ Feynman graph
$\gamma$ in $QED_5$ is 
\begin{equation}
\delta(\gamma)= 5 - \frac{3}{2} E_G - 2 E_F + \frac{1}{2} V
\end{equation}
where $E_G$ and $E_F$ are the number of external gauge and fermion lines,
respectively, and $V$ is the number of vertices.
For the one-loop vacuum polarization tensor, we then have $\delta(\gamma)=
3$, which, taking into account gauge invariance is reduced to $1$.
Moreover, since the divergent terms can only be even polynomials in the
momentum, we are left with a zero degree divergence, i.e., a constant.

Let us now study some immediate properties and consequences that follow
from expressions (\ref{eq:piren1}) and (\ref{eq:piren2}) above. The natural
approach is perhaps to look at its predictions for different momentum
regimes. Let us thus begin by considering the low momentum regime, namely,
$k^2 << m^2$. The leading term, $k^2/m^2 \to 0$ has already been
considered, to impose the renormalization condition $\Pi_R^e \to 0$, which
is not actually a prediction, but rather the fact that the model contains
Coulomb's law at long distances.

The next-to-leading term already contains a non trivial effect. Indeed,  
a simple effect that is sensible to the presence of the flux is
the strength of the Lamb shift, which is determined by
the Darwin term. This can be seen by expanding the renormalized tensor to  
 $(\frac{k}{m})^2$ order in a momentum expansion:
\begin{equation}
\Pi_R(k^2) \;\sim \; -\frac{e^2}{30 \pi} \Big[\frac{m L \,\,\sinh(m
L)}{\cos(\phi)-\cosh(m L)} \big]\, \frac{k^2}{m^2} \;,\; \; k^2 \sim 0 \,, 
\end{equation}
which for the Hydrogen atom produces a corrected potential energy:
\begin{equation}
V_{eff}(r) 
\;=\; - \frac{e^2}{4 \pi r} \,-\,\frac{e^4}{120 \pi^2 m^2} 
\Big[\frac{m L \,\sinh(m L)}{\cosh(m L) - \cos \phi}\Big] \; 
\delta^{(3)}({\mathbf r})\;. 
\end{equation}
The usual correction is obtained when $\phi \to 0$ and $m L \to 0$:
\begin{equation}
V_{eff} (r) \;\to\; - \frac{e^2}{4 \pi r} \,-\,\frac{e^4}{60 \pi^2 m^2} \; 
\delta^{(3)}({\mathbf r}) \;.
\end{equation}
It is interesting to study the shape of the ratio between the corrected and
usual strengths of the respective Darwin terms:
\begin{equation}
\xi(m L, \phi) \;\equiv\;\frac{ 2 m L \,\sinh(m L)}{\cosh(m L) - \cos\phi}\;.  
\end{equation}
The case of a vanishing flux yields simply 
$\xi(m L, 0) \,=\,\frac{\frac{m L}{2}}{\tanh(\frac{m L}{2})}$,  
which for small values of $m L$ approaches $1$, and grows linearly with $m
L$ when \mbox{$m L >> 1$}. 

The opposite regime, when the effect of the flux is maximum, corresponds to
$\phi = \frac{\pi}{2}$:
\begin{equation}
\xi(m L, \frac{\pi}{2}) \;=\; 2 m L \, \tanh(m L)\;.  
\end{equation}
The behaviour in this case is quite different; it tends to zero
quadratically for small $m L$, and also grows linearly in the opposite
case, albeit with a different slope.

Let us now consider the would be large-momentum region for the vacuum
polarization. This regime will be defined by the condition that \mbox{$k^2 >>
m^2$}, although $k$ (and $m$), will be assumed to be much smaller than
$L^{-1}$. The latter is enforced in order to say that the mass of the
Kaluza-Klein modes is much larger than the photon momentum. Under this
assumption, one gets the expression:
\begin{equation}
\Pi^e_R(k^2) \;\sim\; \frac{2\,e^2}{\pi} \int_0^1 d\beta\; \beta(1-\beta)
\, \ln \Big[ 1 +  \beta(1-\beta)\frac{k^2}{m_{eff}^2} \Big] \;,
\end{equation}
where
\begin{equation}
m_{eff} \;\equiv\; \frac{ 2 | \sin \frac{\theta}{2} | }{L} \;.
\end{equation}
We conclude that, as a consequence of the existence of the non-vanishing
flux, the large-momentum behaviour differs from the one that one has in
standard $QED$, by the emergence of an effective mass $m_{eff}$. This mass
should, in order not to spoil the known anti-screening effect at short
distances, be very small. Since $L$ is assumed to be very small, that can
only be achieved with an extremely small $\theta$, namely $\theta << 1$.
Hence,
\begin{equation}
m_{eff} \;\equiv\; \frac{ 2 | \sin \frac{\theta}{2} | }{L} \;\sim\;
\frac{|\theta|}{L} \; << 1 \;.
\end{equation}
In natural units, if $L^{-1} \equiv \Lambda$ is the large momentum scale set
by the Kaluza-Klein modes, and we want $m_{eff}$ to be much smaller than
the electron mass, since only in that situation we recover the expected
behaviour for the effective charge at small distances. Then we should have:
\begin{equation}
|\theta| < < \frac{m}{\Lambda}\;.
\end{equation}

Finally, the parity breaking term, $\Gamma_o$ is simply obtained by taking into account
(\ref{eq:sumj}) and (\ref{eq:defj}):
\begin{equation}
\Gamma_o \,=\, - \frac{i e^2}{16 \pi^2} \,
\Phi \int d^{3+1}x {\tilde F}_{\mu\nu} F_{\mu\nu} \;,
\end{equation}
where we introduced the factor:
\begin{equation}
\Phi \,=\, \sum_{n=-\infty}^\infty \, \phi_n \;;
\end{equation}
the sum of this series is well-known~\cite{Fosco:1997ei}, 
the result being:
\begin{equation}
\Phi \,=\, \arctan \big[ \tanh( \frac{m L}{2} ) \, \tan(\phi/2) \big] \;.  
\end{equation}
The possible effects due to this term are more difficult to elucidate,
since they would require the existence of non-trivial Abelian gauge field
background to manifest themselves. Within the present model, there is
no room to accommodate them, except if singular configurations were

To conclude, we enumerate the main points we wished to convey in this letter: 
by studying the vacuum polarization function in this
$QED$ model with an extra dimension and flux, physical effects due to 
the compactification can be found at the level of the vacuum polarization
tensor. Indeed, the strongest effect is due to the
non-vanishing flux, parametrized by $\theta$, in the large momentum
behaviour of the effective charge. We see that $\theta$ should be much
smaller than the ratio between the electron mass and the (momentum) scale
induced by the inverse of the compactification radius in order for this
effect to be hidden.
Besides, the effect of the non-vanishing flux is maximum when it reaches
$\pi$. This is to be expected, since in that case there is no massless
mode, and then there is no natural way to reduce the theory at the level of
the fermionic field. That is, on the other hand, the case when $\phi=0$,
since it means that the $n=0$ mode finds a natural $3+1$ dimensional
interpretation and there is a smooth limit when $L \to 0$.
Finally, parity breaking effects might be expected, only if there were
a compelling reason to know that the gauge field itself adopts a
topologically non-trivial configuration.

\section*{Acknowledgements}
C.D.F. and C.C.T. thank CONICET for financial
support.  A.P.C.M.  and I.R. thank CNPq/MCT and FAPERJ
for partial financial support.


\end{document}